# Analysis of Selective-Decode and Forward Relaying Protocol Over κ-μ Fading Channel Distribution

Ravi Shankar[1, 2], Lokesh Bhardwaj[1], Ritesh Kumar Mishra[1]

[1] National Institute of Technology Patna, Patna, India
[2] Madanapalle Institute of Technology & Science, Madanapalle, AP, India
ravishankar@mits.ac.in, targetaim31@gmail.com, ritesh@nitp.ac.in

**Abstract:** In this work, we examine the performance of selective-decode and forward (S-DF) relay systems over κ-μ fading channel condition. We discuss about the probability density function (PDF), system model, and cumulative distribution function (CDF) of κ-μ distributed envelope and signal to noise ratio (SNR) and the techniques to generate samples that follow κ-μ distribution. Specifically, we consider the case where the source-to-relay (S→R), relay-to-destination (R→D) and source-to-destination (S→D) link is subject to the independent and identically distributed (*i.i.d.*) κ-μ fading. From the simulation results, the enhancement in the symbol error rate (SER) with a stronger line of sight (LOS) component is observed. This shows that S-DF relaying systems can perform well even in the non-fading or LOS conditions. Monte Carlo simulations are conducted for various values of fading parameters and the outcomes closely match with theoretical outcomes which validate the derivations.
**Keywords:** multiple input multiple output, selective decode and forward, symbol error rate, channel fading, relaying protocol, signal to noise ratio, channel state information.

## 1 Introduction

The 5th generation (5G) wireless communication systems will require a major paradigm shift to meet the increasing demand for reliable connectivity through high data rates, low latency, better energy efficiency, and Femto cell-based relays [1-3]. These relay nodes are equipped with energy harvesting (EH) techniques in the relaying network to improve energy efficiency. Cooperative communication is the natural choice for 5G wireless communication system, and it is adopted in 3rd generation partnership project (3GPP), universal mobile telecommunications service (UMTS), long term evolution (LTE)-Advanced and IEEE 802.11 because the nodes in the cooperative communication network can share their resources with each other during the signal transmission [4-6]. Also, it is incorporated into numerous 5G wireless applications, such as machine-to-machine (M2M), device-to-device (D2D), cognitive radio (CR), high speed terrestrial network (HSTN) & free space optical (FSO) communication [7-10].

Relay-assisted cooperation is the first step towards the 5G system that is expected to deliver up to 20Gbps in downlink (D/L) and 10Gbps in uplink (U/L) will be benchmarked by network operators during initial rollouts in the next few years [11-12]. The relay infrastructure does not need wired network connection, thus offering a reduction in the backhaul costs of the operator. Through the additional cooperative diversity inherent in such wireless systems, cooperative wireless communication significantly improves end-to-end reliability. If the direct source-to-destination (SD) channel is in a deep fade, the main advantage of the cooperative communication is that the destination node can still receive the source signal via the relay node.

According to the signal receiving and transmitting, there are two basic methods of relaying: analog & digital. Analog relaying is also called non-regenerative relaying, in which signals are not required to be digitized before sent by relays. Amplify-and-forward (AF) is a kind of analog relaying. On the other hand, before transmitting them to the destination node, a relay node uses the digital relay protocol to decode and encode signals. Consequently, digital relaying is also known as regenerative relaying. The main drawback of the AF protocol is noise amplification. The relay may transmit the erroneous signal to the destination node in the case of the decode-and-forward (DF) relaying protocol. The S-DF protocol has been proposed to overcome the disadvantage of noise amplification and erroneous decoding related to AF and DF respectively [13-17]. To overcome the problem of noise amplification and relay error propagation, S-DF protocol is used in 5G wireless system. In S-DF relaying network, relay forward only correctly decoded signal otherwise it will remain idle. Moreover, the network connectivity and data transmission rate of S-DF relaying can be further increased by using multiple-input multiple-output (MIMO) in conjunction with space-time block-code (STBC) [18].

In the works [19-20], the authors investigated the S-DF relaying network over Rayleigh flat fading channel. In [19], the authors investigated the pairwise error probability (PEP) performance of the S-DF relaying network over Rayleigh flat fading channel considering the ideal channel conditions. In [21-23], the authors investigated the S-DF relaying network over Nakagami-m fading channel conditions.

In [23], the authors investigated the dual hop (DH) S-DF relaying network over frequency flat Nakagami-m fading channel conditions considering the ideal channel conditions. However, the papers [19-23], did not consider the non-homogeneous fading channel conditions. The performance of wireless communication systems is significantly influenced by stochastic modeling & characterization of the fading links between the communicating nodes. In the development of efficient wireless communication schemes & protocols, accurate stochastic modeling is subsequently very critical. In the literature, a variety of stochastic/statistical distributions have been developed to model the small-scale fluctuations in the transmitted signal envelope over fading channels, such as Nakagami-m, Rayleigh and Weibull [24-26]. None of the stochastic model described, however, captures the non-linearity of the medium of propagation. The $\kappa$-$\mu$ distribution is suitable choice for LOS applications and on the other hand, for non-LOS, $\eta$–$\mu$ fading distribution is better suited. In the work [27], the authors proposed $\eta$–$\mu$ & $\kappa$-$\mu$ non-homogeneous fading distributions for LOS & non-LOS, components, respectively. In [28-33] the authors investigated the relaying network over $\kappa$-$\mu$ & $\eta$–$\mu$ fading channel conditions. In [34], the authors investigated the SER performance of the DF relaying network over $\eta$–$\mu$ & $\kappa$-$\mu$ fading channel conditions. The exact SER expression is derived for the M-ary phase shift keying (PSK) modulated schemes.

In this work, SER expressions is obtained for S-DF relaying network and additional diversity gain is achieved due to use of MIMO in conjunction with STBC. We consider the κ-μ fading distribution, so it is well suited for LOS applications.

The work is organized as follows: In section 2, SER is investigated over $\kappa$-$\mu$ fading channel conditions. In this section closed form SER expression is derived using moment generating function (MGF) based approach. In section 3, simulation results are given and in section 4 conclusion is given.

## 2  End-to-end Symbol Error Analysis

Consider MIMO- STBC S-DF relaying network with $K$ number of relay nodes. In all the analysis associated with SER performance analysis of S-DF relaying network, we will assume $N_S \times N_R$ MIMO systems. Where $N_S$ are the number of antennas equipped at the source node and $N_R$ are the number of antennas equipped at the relay nodes. Since both source and relay nodes use the same orthogonal STBC code, we take $N_S = N_R = N$. It is presumed that orthogonal STBC code is conveyed over $T$ time slots. So, an orthogonal STBC codeword for complete STBC communication can be agreed by a matrix with dimensions $N_S \times T$. It was before it was argued that orthogonal STBC codewords can be managed in different aerials and then processed data can be combined together to get an effective data, which is similar to the maximum ratio combiner (MRC) [35]. The orthogonal STBC is designed such that the vectors representing any pair of columns taken from the source-to-$r^{th}$ relay coding matrix $H_{sr}$ is orthogonal, i.e., the STBC converts the vector channel in scalar channel. For orthogonal STBC designs the conditional SNR at the receiver can be given as Euclidean norm or the Frobenius norm of the channel times average SNR as in (1) [4].

$$\gamma_{sr} = \overline{\gamma}_{sr} \left\| H_{sr} \right\|_F^2, \tag{1}$$

where $\gamma_{sr}$ denotes the conditional SNR of the source-to-$r^{th}$ relay fading link, $\overline{\gamma}_{sr}$ denotes the average conditional SNR of the source-to-$r^{th}$ relay fading link and $\left\| H_{sr} \right\|_F^2$ denotes the Frobenius norm or $L_2$ norm. Every element of $H_{sr}$ is i.i.d. $\kappa$-$\mu$ distributed random variables (RVs). The suggestion of $\kappa$-$\mu$ channel fading was given in [36] as a generalized distribution to model non-homogeneous fading environment. Like $\eta$–$\mu$ & Nakagami-m fading channel distribution, it has been observed that, for $\kappa$-$\mu$ fading channel conditions, the multipath components form clusters. Each cluster has several scattered multipath components. The delay spread of different clusters is relatively larger than the delay spread of multipath components within a cluster. Every cluster is assumed to have the same average power. Unlike in $\eta$–$\mu$ fading and like Nakagami-m fading, it is assumed that the in-phase and quadrature phase components are independent and have equal powers in $\kappa$-$\mu$ fading. However, each cluster is assumed to have some dominant components considered to be LOS components. In such a model, the representation of the envelope $\beta$ of the fading signal is slightly different from that of Nakagami-m and/or η–μ fading. It can be given as [36-39],

$$\beta^2 = \sum_{j=0}^{J}((I_j + \phi_j)^2 + (\Omega_j + \alpha_j)^2), \tag{2}$$

where $J$ is the number of clusters in the received signal, and $(I_j + \phi_j)$ & $(\Omega_j + \alpha_j)$ are respectively the in-phase and quadrature phase component of the resultant signal of the $j^{th}$ cluster. Both $I_j$ & $\Omega_j$ are mutually independent and zero-mean circular-shift complex Gaussian (ZMCSCG), i.e., $E[I_j] = E[\Omega_j] = 0$ & equal variance, i.e., $E[I_j^2] = E[\Omega_j^2] = \sigma^2$. $\phi_j$ and $\alpha_j$ denote the in-phase and quadrature components, respectively. The non-zero mean of in-phase and quadrature phase components reveals the presence of a dominant component in the clusters of the received signal. Again, as in the case of Nakagami-m and $\eta$–$\mu$ fading model, the fading amplitude can be expressed as,

$$\beta^2 = \sum_{j=0}^{J} (\mathbb{C}_j^2), \tag{3}$$

where, $\mathbb{C}_j^2 = (I_j + \phi_j)^2 + (\Omega_j + \alpha_j)^2.$ (4)

From the fact that $I_j$ & $\Omega_j$ are Gaussian distributed, it is to be noted here that $\mathbb{C}_j^2$ follows non-central Chi-squared distribution. Unlike Nakagami-*m* & $\eta-\mu$ channel models, *κ-μ* distribution is suitable for model LOS environments. PDF of SNR for an STBC MIMO system over *κ-μ* fading channels can be given by [36-39],

$$p_{\gamma_{sr}}(\gamma_{sr}) = \frac{\mu_{sr} N_S N_R (1+\kappa_{sr})^{\frac{\mu_{sr} N_S N_R + 1}{2}}}{(\kappa_{sr})^{\frac{\mu_{sr} N_S N_R - 1}{2}} e^{\mu_{sr} N_S N_R \kappa_{sr}} \bar{\gamma}_{sr}^{\frac{\mu_{sr} N_S N_R + 1}{2}}} \times \gamma_{sr}^{\frac{\mu_{sr} N_S N_R - 1}{2}} e^{-\frac{\mu_{sr} N_S N_R (1+\kappa_{sr}) \gamma_{sr}}{\bar{\gamma}_{sr}}} I_{\mu_{sr} N_S N_R - 1}\left(2\mu_{sr} N_S N_R \sqrt{\frac{\kappa_{sr}(1+\kappa_{sr})\gamma_{sr}}{\bar{\gamma}_{sr}}}\right), \tag{5}$$

where $\mu_{sr} > 0$ is the channel fading parameter directly related to the number of clusters, $\kappa_{sr}$ denotes the ratio of power in the LOS components to that of scattered components. Note that $\bar{\gamma}_{sr}$ is the expected SNR and is used as a scaling factor of $\|H_{sr}\|_F^2$ in existing literatures to indicate average SNR at the receiver. The instantaneous SER $P_E^{S \to R}(\gamma_{sr})$ of the source-to-$r^{th}$ relay fading link can be expressed as [40-41],

$$P_E^{S \to R}(\gamma_{sr}) = aQ(\sqrt{b\gamma_{sr}}) - cQ^2(\sqrt{b\gamma_{sr}}), \tag{6}$$

where $a$, $b$ & $c$ are modulation dependent parameters listed in Table 1 and $Q(.)$ represents the Gaussian $Q$ function which gives the area under the tail of a Gaussian curve and is defined as [40-41],

$$Q(x) = \frac{1}{\sqrt{2\pi}} \int_x^\infty \exp(-u^2/2) du = \frac{1}{\sqrt{\pi}} \int_x^\infty \exp(-z^2) dz = \frac{erfc(x)}{2}, \tag{7}$$

where $erfc(x)$ is the complementary error function, which is accessible among others in MATLAB software.

**Table 1.** Modulation parameters for various modulation schemes [34-35]

| Modulation Scheme | $a$ | $b$ | $c$ |
|---|---|---|---|
| Binary PSK | 1 | 2 | 0 |
| Binary Frequency Shift Keying | 1 | 1 | 0 |
| M*ary*-PSK | 2 | $2\sin^2\left(\frac{\pi}{M}\right)$ | 0 |
| M*ary*-Pulse Amplitude Modulation | $2\frac{M-1}{M}$ | $\frac{6}{M^2-1}$ | 0 |
| Quadrature PSK | 2 | 2 | 1 |
| Coherent Differential PSK | 2 | 2 | 2 |
| M*ary*-Quadrature Amplitude Modulation (QAM) | $4\frac{\sqrt{M}-1}{\sqrt{M}}$ | $\frac{3}{M-1}$ | $4\left(\frac{\sqrt{M}-1}{\sqrt{M}}\right)^2$ |

The expected SER $\overline{P_E^{S \to R}}$ can be obtained by taking expectation of the instantaneous SER over the PDF of receiving instantaneous SNR. For averaging the conditional SER, we will use the MGF based approach in this section. It can be expressed as [40],

$$\overline{P_E^{S \to R}} = \underbrace{\frac{a}{\pi}\int_0^{\frac{\pi}{2}} M_{\gamma_{sr}}\left(\frac{b}{2\sin^2\theta}\right)d\theta}_{I_1} - \underbrace{\frac{c}{\pi}\int_0^{\frac{\pi}{4}} M_{\gamma_{sr}}\left(\frac{b}{2\sin^2\theta}\right)d\theta}_{I_2}, \quad (8)$$

where $M_{\gamma_{sr}}(.)$ is the MGF of received conditional SNR. MGF of $\kappa$-$\mu$ distributed instantaneous SNR is given as [40-41],

$$M_{\gamma_{sr}}(s) = \int_0^\infty p_{\gamma_{sr}}(\gamma_{sr})e^{-s\gamma_{sr}}d\gamma_{sr} = \left(\frac{\mu_{sr}N_S N_R(1+\kappa_{sr})}{\mu_{sr}N_S N_R(1+\kappa_{sr})+s\overline{\gamma}_{sr}}\right)^{\mu_{sr}N_S N_R} e^{\frac{(\mu_{sr}N_S N_R)^2 \kappa_{sr}(1+\kappa_{sr})}{\mu_{sr}N_S N_R(1+\kappa_{sr})+s\overline{\gamma}_{sr}}-\kappa_{sr}\mu_{sr}N_S N_R}. \quad (9)$$

$I_1$ can be expressed in terms of confluent Hypergeometric function [42], as evaluated in Appendix A.

$I_1 =$

$$\frac{a}{\pi}\frac{\sqrt{b\overline{\gamma}_{sr}}(\mu_{sr}N_S N_R \kappa_{sr})^{\mu_{sr}N_S N_R}}{\sqrt{2\pi\mu_{sr}N_S N_R(1+\kappa_{sr})}}\left(\frac{2\mu_{sr}N_S N_R(1+\kappa_{sr})}{2\mu_{sr}N_S N_R(1+\kappa_{sr})+b\overline{\gamma}_{sr}}\right)^{\mu_{sr}N_S N_R+1/2}\frac{\Gamma\left(\mu_{sr}N_S N_R+\frac{1}{2}\right)\sqrt{\pi}}{\Gamma(\mu_{sr}N_S N_R+1)} \times$$

$$\sum_{J=0}^\infty \sum_{n=0}^\infty \frac{\left(\mu_{sr}N_S N_R+\frac{1}{2}\right)_{J+n}(1)_J \left(\frac{2\mu_{sr}N_S N_R(1+\kappa_{sr})}{2\mu_{sr}N_S N_R(1+\kappa_{sr})+b\overline{\gamma}_{sr}}\right)^J \left(\frac{2\mu_{sr}N_S N_R(1+\kappa_{sr})}{2\mu_{sr}N_S N_R(1+\kappa_{sr})+b\overline{\gamma}_{sr}}\right)^n}{(\mu_{sr}N_S N_R+1)_{J+n} J!n!}.$$

(10)

where $\Gamma(x)$ represents the Gamma function [42] and $(x)_n$ denotes the descending factorial [44-45], expressed as, $(x)_n = \frac{\Gamma(x+1)}{\Gamma(x-n+1)}$. Also, $I_2$ can be expressed in terms of confluent Lauricella's Hypergeometric function [43], as evaluated in Appendix B.

$$I_2 = \frac{c\sqrt{b\overline{\gamma}_{sr}}(\mu_{sr}N_S N_R \kappa_{sr})^{\mu_{sr}N_S N_R}}{2\pi\sqrt{2\mu_{sr}N_S N_R(1+\kappa_{sr})}}\left(\frac{\mu_{sr}N_S N_R(1+\kappa_{sr})}{\mu_{sr}N_S N_R(1+\kappa_{sr})+b\overline{\gamma}_{sr}}\right)^{\mu_{sr}N_S N_R+\frac{1}{2}}\frac{\Gamma\left(\mu_{sr}N_S N_R+\frac{1}{2}\right)\sqrt{\pi}}{\Gamma(\mu_{sr}N_S N_R+1)} \times$$

$$\sum_{J=0}^\infty \sum_{n=0}^\infty \sum_{p=0}^\infty \frac{\left\{\begin{array}{l}\left(\mu_{sr}N_S N_R+\frac{1}{2}\right)_{J+n+p}(1)_m\left(\frac{1}{2}\right)_n\left(\frac{\mu_{sr}N_S N_R(1+\kappa_{sr})}{\mu_{sr}N_S N_R(1+\kappa_{sr})+b\overline{\gamma}_{sr}}\right)^J \times \\ \left(\frac{2\mu_{sr}N_S N_R(1+\kappa_{sr})}{2\mu_{sr}N_S N_R(1+\kappa_{sr})+2b\overline{\gamma}_{sr}}\right)^n \left(\frac{\mu_{sr}N_S N_R(1+\kappa_{sr})}{\mu_{sr}N_S N_R(1+\kappa_{sr})+b\overline{\gamma}_{sr}}\right)^n\end{array}\right\}}{\left(\frac{3}{2}\right)_{J+n+p} J!n!p!}.$$

(11)

Following the similar analysis, the SER for the S→D fading link can be expressed below [40],

$$\overline{P_E^{S \to D}} = \underbrace{\frac{a}{\pi}\int_0^{\frac{\pi}{2}} M_{\gamma_{sd}}\left(\frac{b}{2\sin^2\theta}\right)d\theta}_{K_1} - \underbrace{\frac{c}{\pi}\int_0^{\frac{\pi}{4}} M_{\gamma_{sd}}\left(\frac{b}{2\sin^2\theta}\right)d\theta}_{K_2}, \quad (12)$$

where,

$$K_1 = \frac{a}{\pi}\frac{\sqrt{b\overline{\gamma}_{sd}}(\mu_{SD}N_S N_D \kappa_{SD})^{\mu_{SD}N_S N_D}}{\sqrt{2\pi\mu_{SD}N_S N_D(1+\kappa_{SD})}}\left(\frac{2\mu_{SD}N_S N_D(1+\kappa_{SD})}{2\mu_{SD}N_S N_D(1+\kappa_{SD})+b\overline{\gamma}_{sd}}\right)^{\mu_{SD}N_S N_D+1/2}\frac{\Gamma\left(\mu_{SD}N_S N_D+\frac{1}{2}\right)\sqrt{\pi}}{\Gamma(\mu_{SD}N_S N_D+1)} \times$$

$$\sum_{J=0}^\infty \sum_{n=0}^\infty \frac{\left(\mu_{SD}N_S N_D+\frac{1}{2}\right)_{J+n}(1)_J \left(\frac{2\mu_{SD}N_S N_D(1+\kappa_{SD})}{2\mu_{SD}N_S N_D(1+\kappa_{SD})+b\overline{\gamma}_{sd}}\right)^J \left(\frac{2\mu_{SD}N_S N_D(1+\kappa_{SD})}{2\mu_{SD}N_S N_D(1+\kappa_{SD})+b\overline{\gamma}_{sd}}\right)^n}{(\mu_{SD}N_S N_D+1)_{J+n} J!n!},$$

(13)

where $\mu_{SD} > 0$ is the channel fading parameter directly related to the number of clusters and $\kappa_{SD}$ denotes the ratio of power in the LOS components to that of scattered components for S→D fading links. $\overline{\gamma}_{sd}$ denotes the average SNR of the source-to-destination fading link.

$$K_2 = \frac{c}{\pi} \frac{\sqrt{b\bar{\gamma}_{sd}} (\mu_{SD} N_S N_D \kappa_{SD})^{\mu_{SD} N_S N_D}}{2\sqrt{2\mu_{SD} N_S N_D (1+\kappa_{SD})}} \left( \frac{\mu_{SD} N_S N_D (1+\kappa_{SD})}{\mu_{SD} N_S N_D (1+\kappa_{SD}) + b\bar{\gamma}_{sd}} \right)^{\mu_{SD} N_S N_D + \frac{1}{2}} \frac{\Gamma\left(\mu_{SD} N_S N_D + \frac{1}{2}\right)\sqrt{\pi}}{\Gamma(\mu_{SD} N_S N_D + 1)} \times$$

$$\sum_{J=0}^{\infty}\sum_{n=0}^{\infty}\sum_{p=0}^{\infty} \frac{\left\{\left(\mu_{SD} N_S N_D + \frac{1}{2}\right)_{J+n+p} (1)_J \left(\frac{1}{2}\right)_n \left(\frac{\mu_{SD} N_S N_D (1+\kappa_{SD})}{\mu_{SD} N_S N_D (1+\kappa_{SD}) + b\bar{\gamma}_{sd}}\right)^J \times \left(\frac{2\mu_{SD} N_S N_D (1+\kappa_{SD})}{2\mu_{SD} N_S N_D (1+\kappa_{SD}) + 2b\bar{\gamma}_{sd}}\right)^n \left(\frac{\mu_{SD} N_S N_D (1+\kappa_{SD})}{\mu_{SD} N_S N_D (1+\kappa_{SD}) + b\bar{\gamma}_{sd}}\right)^n \right\}}{\left(\frac{3}{2}\right)_{J+n+p} J!n!p!}. \quad (14)$$

Also, the SER for the R→D fading link can be expressed below [40],

$$\overline{P_E^{R \to D}} = \underbrace{\frac{a}{\pi} \int_0^{\frac{\pi}{2}} M_{\gamma_{rd}}\left(\frac{b}{2\sin^2\theta}\right) d\theta}_{\psi_1} - \underbrace{\frac{c}{\pi} \int_0^{\frac{\pi}{4}} M_{\gamma_{rd}}\left(\frac{b}{2\sin^2\theta}\right) d\theta}_{\psi_2}, \quad (15)$$

where,

$$\psi_1 = \frac{a}{\pi} \frac{\sqrt{b\bar{\gamma}_{rd}} (\mu_{RD} N_R N_D \kappa_{RD})^{\mu_{RD} N_R N_D}}{\sqrt{2\mu_{RD} N_R N_D (1+\kappa_{RD})}} \left( \frac{2\mu_{RD} N_R N_D (1+\kappa_{RD})}{2\mu_{RD} N_R N_D (1+\kappa_{RD}) + b\bar{\gamma}_{rd}} \right)^{\mu_{RD} N_R N_D + 1/2} \frac{\Gamma\left(\mu_{RD} N_R N_D + \frac{1}{2}\right)\sqrt{\pi}}{\Gamma(\mu_{RD} N_R N_D + 1)} \times$$

$$\sum_{J=0}^{\infty}\sum_{n=0}^{\infty} \frac{\left(\mu_{RD} N_R N_D + \frac{1}{2}\right)_{J+n} (1)_m \left(\frac{2\mu_{RD} N_R N_D (1+\kappa_{RD})}{2\mu_{RD} N_R N_D (1+\kappa_{RD}) + b\bar{\gamma}_{rd}}\right)^J \left(\frac{2\mu_{RD} N_R N_D (1+\kappa_{RD})}{2\mu_{RD} N_R N_D (1+\kappa_{RD}) + b\bar{\gamma}_{rd}}\right)^n}{(\mu_{RD} N_R N_D + 1)_{J+n} J!n!}. \quad (16)$$

$$\psi_2 = \frac{c}{\pi} \frac{\sqrt{b\bar{\gamma}_{rd}} (\mu_{RD} N_R N_D \kappa_{RD})^{\mu_{RD} N_R N_D}}{2\sqrt{2\mu_{RD} N_R N_D (1+\kappa_{RD})}} \left( \frac{\mu_{RD} N_R N_D (1+\kappa_{RD})}{\mu_{RD} N_R N_D (1+\kappa_{RD}) + b\bar{\gamma}_{rd}} \right)^{\mu_{RD} N_R N_D + \frac{1}{2}} \frac{\Gamma\left(\mu_{RD} N_R N_D + \frac{1}{2}\right)\sqrt{\pi}}{\Gamma(\mu_{RD} N_R N_D + 1)} \times$$

$$\sum_{J=0}^{\infty}\sum_{n=0}^{\infty}\sum_{p=0}^{\infty} \frac{\left\{\left(\mu_{RD} N_R N_D + \frac{1}{2}\right)_{J+n+p} (1)_m \left(\frac{1}{2}\right)_n \left(\frac{\mu_{RD} N_R N_D (1+\kappa_{RD})}{\mu_{RD} N_R N_D (1+\kappa_{RD}) + b\bar{\gamma}_{rd}}\right)^J \times \left(\frac{2\mu_{RD} N_R N_D (1+\kappa_{RD})}{2\mu_{RD} N_R N_D (1+\kappa_{RD}) + 2b\bar{\gamma}_{rd}}\right)^n \left(\frac{\mu_{RD} N_R N_D (1+\kappa_{RD})}{\mu_{RD} N_R N_D (1+\kappa_{RD}) + b\bar{\gamma}_{rd}}\right)^n \right\}}{\left(\frac{3}{2}\right)_{J+n+p} J!n!p!}, \quad (17)$$

where $\mu_{RD} > 0$ is the channel fading parameter directly related to the number of clusters and $\kappa_{RD}$ denotes the ratio of power in the LOS components to that of scattered components for R→D fading links and $\bar{\gamma}_{rd}$ denotes the average SNR of the source-to-destination fading link. The error probability of the cooperation mode, $\overline{P_E^{S \to D, R \to D}}$ can be expressed as [41], [44],

$$\overline{P_E^{S \to D, R \to D}} = \left\{ \underbrace{\frac{a}{\pi} \int_0^{\frac{\pi}{2}} M_{\gamma_{sd}}\left(\frac{b}{2\sin^2\theta}\right) d\theta}_{I_1} - \underbrace{\frac{c}{\pi} \int_0^{\frac{\pi}{4}} M_{\gamma_{sd}}\left(\frac{b}{2\sin^2\theta}\right) d\theta}_{I_2} \right\} \times$$

$$\left\{ \underbrace{\frac{a}{\pi} \int_0^{\frac{\pi}{2}} M_{\gamma_{rd}}\left(\frac{b}{2\sin^2\theta}\right) d\theta}_{\psi_1} - \underbrace{\frac{c}{\pi} \int_0^{\frac{\pi}{4}} M_{\gamma_{rd}}\left(\frac{b}{2\sin^2\theta}\right) d\theta}_{\psi_2} \right\}. \quad (18)$$

Where $\overline{P_E^{S \to D, R \to D}}$ represents the cooperation mode of signal transmission. In the relaying phase if the relay decodes correctly then the destination gets signal from the relay node as well as from the source node. The optimal combining is done at the destination node using the maximal ratio combining schemes. The end-to-end SER of the cooperative communication fading link can be expressed as,

$$\overline{P_E} = \overline{P_E^{S \to R}} \times \overline{P_E^{S \to D}} + \left(1 - \overline{P_E^{S \to R}}\right) \times \overline{P_E^{S \to D, R \to D}}. \tag{19}$$

The end-to-end SER of the cooperative communication fading link can be obtained by substituting (8) (12) and (18) into (19).

## 3  Simulation Results

For a MIMO-STBC S-DF relaying network, we demonstrate simulation plots of the average SER over non-homogeneous fading channel conditions. Monte Carlo simulations are conducted, and matrix laboratory (MATLAB) software has been used for simulations. In Figs. 1-3, for simplicity reasons we take $\mu_{RD} = \mu_{SD} = \mu_{SR} = \mu$ and $\kappa_{RD} = \kappa_{SD} = \kappa_{SR} = \kappa$. The theoretical expressions get in $\kappa$-$\mu$ fading is in the infinite series form; however, these series converge very rapidly with an increase in the number of summations terms (*N*), e.g., *N* = 15 is enough to attain accuracy up 4 decimal numbers. For better and assured precision, the corresponding analysis is performed with *N*=20. In Fig. 1, we considered the equal power allocation factors with Q-PSK modulated symbols, the average end-to-end error probability versus SNR plots are shown with clear detection over fading channels of $\kappa - \mu$. The average SER is plotted for *μ*=1, and varying *κ* using (16). We observe that the increment in performance is more for increase in *κ*. In Fig.2, SER vs. SNR in dB is plotted for various values of *κ* and for fixed value of *μ*. In Fig. 3, SER vs. SNR in dB plot is given for various values of *μ* and for fixed value of *κ*. It has been shown that with increase the value of *μ*, the SER performance improves.

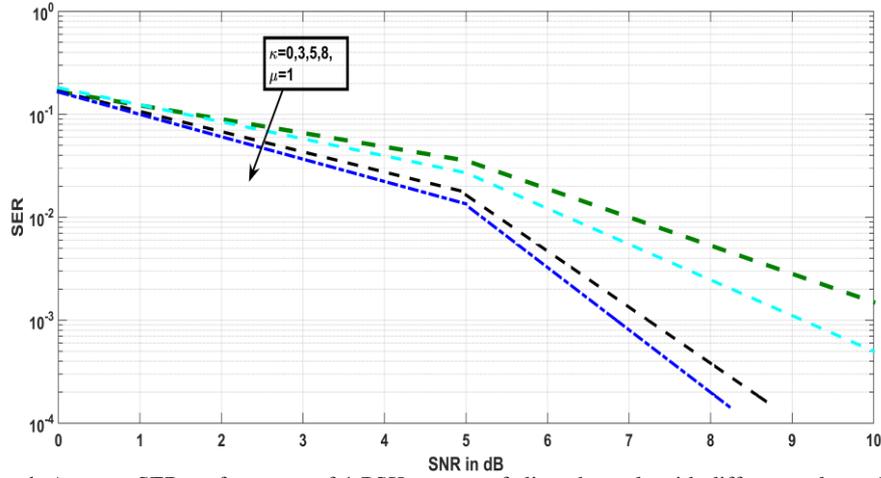

**Fig. 1.** Average SER performance of 4-PSK over *κ-μ* fading channels with different values of *κ*.

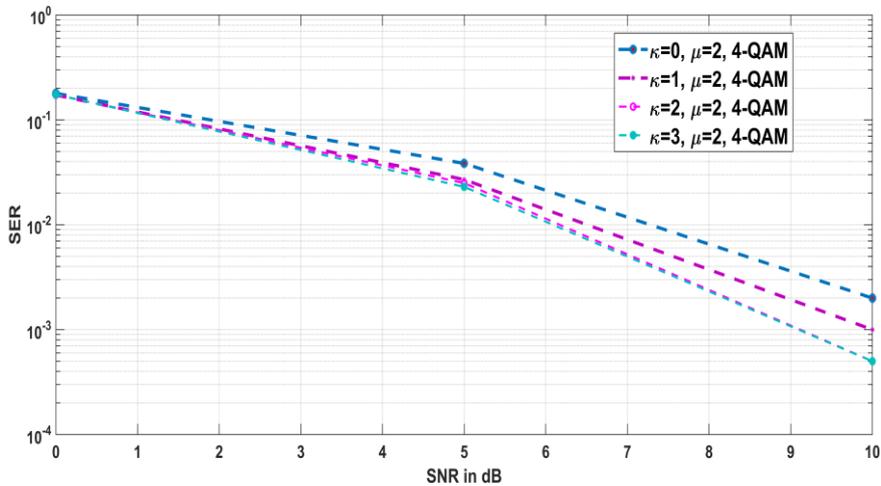

**Fig. 2.** Average SER performance of 4-QAM over κ-μ fading channels with different values of κ.

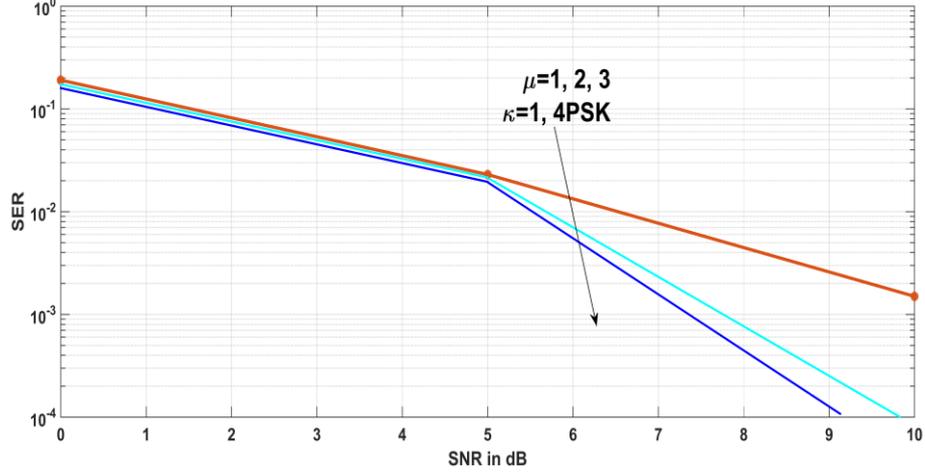

**Fig. 3.** Average SER performance of 4-PSK over κ-μ fading channels with different values of μ.

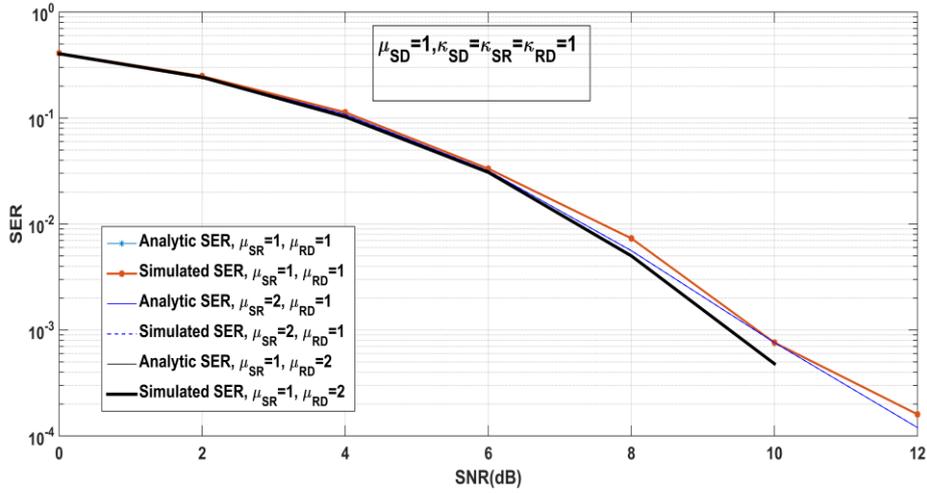

**Fig. 4.** SER vs. SNR in dB plots of 4-QAM over κ-μ links for various values of $\mu_{RD}$ & $\mu_{SR}$ & $\mu_{SD}=1, \kappa_{SD}=\kappa_{SR}=\kappa_{RD}=1$.

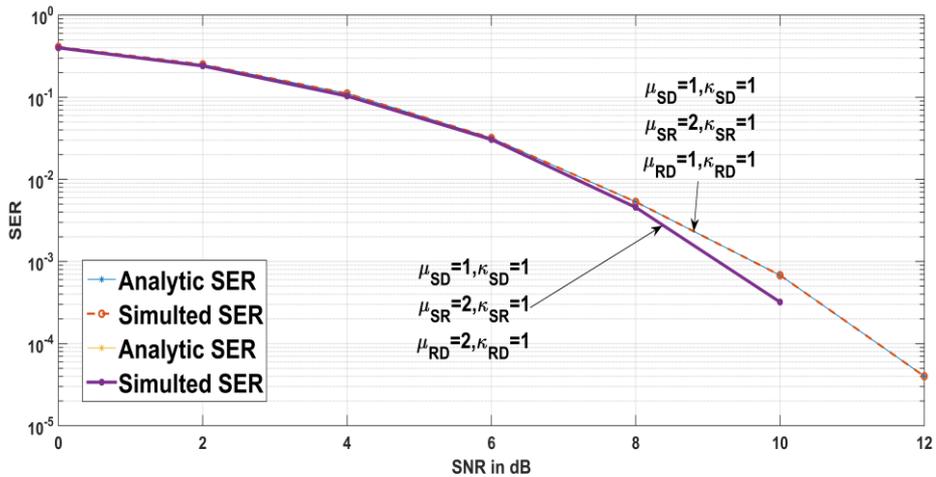

**Fig. 5**. SER vs. SNR in dB plots of 4-QAM over κ-μ links for various values of $\kappa_{ij}$ & $\mu_{ij}$.

The SER vs. SNR in dB is demonstrated in Figs. 4-5 for 4-QAM modulated symbols over $\kappa-\mu$ links. In Fig. 4, the SER is plotted for $\mu_{SD}=1, \kappa_{SD}=\kappa_{SR}=\kappa_{RD}=1$, & varying $\mu_{SR}$ & $\mu_{RD}$ applying (16). There is an important finding that the increase in SER performance is more likely to increase in $\mu_{RD}$ than in $\mu_{SR}$. In Fig. 5, $\kappa_{ij}$ & $\mu_{ij}$; $(ij \in SD, SR, RD)$ are fixed for S→R & S→D link, and varied for R→D link. It is significant to note that increase in $\kappa_{RD}$ has lesser impact on performance than increase in $\mu_{RD}$. The S→R and R→D links with greater values of fading parameters $\mu$ & $\kappa$ shows better performance and among them the parameters of R→D link dominate. For contrast, the increase in $\mu_{RD}$ is greater than the increase in $\kappa_{RD}$.

## 4 Conclusion

We have investigated CF expressions of the average SER for a MIMO STBC S-DF relaying network over κ-μ faded links when input is Q-PSK and 4-QAM modulated. The average SER of QAM and QPSK are presented in the drawings. Specifically, we consider the case where the S→R, R→D and S→D link is subject to the *i.i.d.* κ-μ fading. From simulation results, the enhancement in SER with a stronger LOS component is observed.

## Appendix A

*Solution of* $I_1 = \dfrac{a}{\pi} \displaystyle\int_0^{\frac{\pi}{2}} M_{\gamma_{sr}}\left(\dfrac{b}{2\sin^2\theta}\right) d\theta$:

Let, $t = \dfrac{\mu_{sr}(1+\kappa_{sr})}{\mu_{sr}(1+\kappa_{sr}) + \dfrac{b\bar{\gamma}_{sr}}{2\sin^2(\theta)}}$, (A.1)

After performing some mathematical manipulations, $\sin^2(\theta)$ is expressed as,

$\sin^2(\theta) = \dfrac{b\bar{\gamma}_{sr} t}{2\mu_{sr}(1+\kappa_{sr})(1-t)}$, (A.2)

$\sin(\theta) = \sqrt{\dfrac{b\bar{\gamma}_{sr} t}{2\mu_{sr}(1+\kappa_{sr})(1-t)}}$, (A.3)

$\cos(\theta) = \sqrt{1 - \dfrac{b\bar{\gamma}_{sr} t}{2\mu_{sr}(1+\kappa_{sr})(1-t)}}$, (A.4)

After differentiating (A.2) with respect to $dt$, we get,

$2\sin(\theta)\cos(\theta) d\theta = \dfrac{b\bar{\gamma}_{sr} dt}{2\mu_{sr}(1+\kappa_{sr})(1-t)^2}$, (A.5)

The lower and upper limits of integral $I_1$ changes from 0 to 0 & from $\pi/2$ to $\dfrac{\mu_{sr}(1+\kappa_{sr})}{\mu_{sr}(1+\kappa_{sr}) + b\bar{\gamma}_{sr}/2}$.

Substituting (A.2) - (A.5) into integral $I_1$, we get,

$I_1 = \dfrac{a}{\pi} \displaystyle\int_0^{\frac{\mu_{sr}(1+\kappa_{sr})}{\mu_{sr}(1+\kappa_{sr})+b\bar{\gamma}_{sr}/2}} \dfrac{b\bar{\gamma}_{sr}(\mu_{sr}\kappa_{sr}t)^{\mu_{sr}} \exp(t)(1-t)^{-1}}{2\sqrt{b\bar{\gamma}_{sr} t}\sqrt{2\mu_{sr}(1+\kappa_{sr})(1-t) - b\bar{\gamma}_{sr} t}} dt$, (A.6)

For further simplification of this integral, it can be brought in the form of confluent hypergeometric function with the substitution,

$y = \dfrac{2\mu_{sr}(1+\kappa_{sr}) + b\bar{\gamma}_{sr}}{2\mu_{sr}(1+\kappa_{sr})} t$, (A.7)

The above substitution converts the upper limit of the integral to unity without changing the lower limit of the integration. This makes it easy to represent this integral into standard form of confluent hypergeometric function of two variables. Confluent hypergeometric function is defined as,

$\Phi_1(a,b,c,x,y) = \dfrac{\Gamma(c)}{\Gamma(a)\Gamma(c-a)} \displaystyle\int_0^1 t^{a-1}(1-t)^{c-a-1}(1-xt)^{-b}\exp(yt) dt$, (A.8)

Further simplifications after substitution of (A.7) into (A.6) bring the integral in the form that can be given as,

$$I_1 = \frac{a}{\pi}\int_0^1 \frac{b\bar{\gamma}_{sr} y^{\mu-1/2} \exp\left(\frac{2\mu_{sr}(1+\kappa_{sr})}{2\mu_{sr}(1+\kappa_{sr})+b\bar{\gamma}_{sr}}y\right)\left(1-\frac{2\mu_{sr}(1+\kappa_{sr})}{2\mu_{sr}(1+\kappa_{sr})+b\bar{\gamma}_{sr}}y\right)^{-1}}{2\sqrt{b\bar{\gamma}_{sr}}\sqrt{2\mu_{sr}(1+\kappa_{sr})(1-y)}}\left(\frac{2\mu_{sr}(1+\kappa_{sr})}{2\mu_{sr}(1+\kappa_{sr})+b\bar{\gamma}_{sr}}\right)dy,$$
(A.9)

$$= \frac{a}{\pi}\frac{\sqrt{b\bar{\gamma}_{sr}}(\mu_{sr}\kappa_{sr})^{\mu_{sr}}}{2}\left(\frac{2\mu_{sr}(1+\kappa_{sr})}{2\mu_{sr}(1+\kappa_{sr})+b\bar{\gamma}_{sr}}\right)^{\mu_{sr}+1/2}\int_0^1 \frac{b\bar{\gamma}_{sr} y^{\mu_{sr}-1/2}\exp\left(\frac{2\mu_{sr}(1+\kappa_{sr})}{2\mu_{sr}(1+\kappa_{sr})+b\bar{\gamma}_{sr}}y\right)(1-y)^{-1/2}}{\sqrt{2\mu_{sr}(1+\kappa_{sr})(1-y)}\left(1-\frac{2\mu_{sr}(1+\kappa_{sr})}{2\mu_{sr}(1+\kappa_{sr})+b\bar{\gamma}_{sr}}y\right)}dy,$$
(A.10)

The above expression can be compared with the definition of confluent hypergeometric function given in (A.8) to obtain the arguments of the function as,

$$a = \mu_{sr} + 1/2,$$ (A.11)
$$b = 1,$$ (A.12)
$$c = \mu_{sr} + 1,$$ (A.13)
$$x = \frac{2\mu_{sr}(1+\kappa_{sr})}{2\mu_{sr}(1+\kappa_{sr})+b\bar{\gamma}_{sr}},$$ (A.14)
$$y = \frac{2\mu_{sr}(1+\kappa_{sr})}{2\mu_{sr}(1+\kappa_{sr})+b\bar{\gamma}_{sr}},$$ (A.15)

Thus, $I_1$ can be finally evaluated in the form of confluent hypergeometric function as,

$$I_1 =$$

$$\frac{a}{\pi}\frac{\sqrt{b\bar{\gamma}_{sr}}(\mu_{sr}\kappa_{sr})^{\mu_{sr}}}{2\sqrt{2\mu_{sr}(1+\kappa_{sr})}}\left(\frac{2\mu_{sr}(1+\kappa_{sr})}{2\mu_{sr}(1+\kappa_{sr})+b\bar{\gamma}_{sr}}\right)^{\mu_{sr}+1/2}\frac{\Gamma(\mu_{sr}+\frac{1}{2})\sqrt{\pi}}{\Gamma(\mu_{sr}+1)}\times$$
$$\Phi_1\left(\mu_{sr}+\frac{1}{2},1,\mu_{sr}+1,\frac{2\mu_{sr}(1+\kappa_{sr})}{2\mu_{sr}(1+\kappa_{sr})+b\bar{\gamma}_{sr}},\frac{2\mu_{sr}(1+\kappa_{sr})}{2\mu_{sr}(1+\kappa_{sr})+b\bar{\gamma}_{sr}}\right),$$
(A.16)

Next, we will discuss the solution of $I_2$.

**Appendix B**

One can proceed for solution of I2 following the steps used for solution of I1 earlier. We take the same substitution as that for I1 in (A.7). Thus, the same expressions will be used in the integral as discussed in equations (A.11)-(A.15). However, the upper limit of the integral will now be $\frac{\mu_{sr}(1+\kappa_{sr})}{\mu_{sr}(1+\kappa_{sr})+b\bar{\gamma}_{sr}}$. Now, it can be represented as,

$$I_2 = \frac{c}{\pi}\int_0^{\frac{\mu_{sr}(1+\kappa_{sr})}{\mu_{sr}(1+\kappa_{sr})+b\bar{\gamma}_{sr}}}\frac{b\bar{\gamma}_{sr}(\mu_{sr}\kappa_{sr}t)^{\mu_{sr}}\exp(t)(1-t)^{-1}}{2\sqrt{b\bar{\gamma}_{sr}t}\sqrt{2\mu_{sr}(1+\kappa_{sr})(1-t)-b\bar{\gamma}_{sr}t}}dt,$$ (A.17)

For further simplification of this integral, it can be brought in the form of
confluent Lauricella's hypergeometric function of three variables with the substitution

$$y = \frac{\mu_{sr}(1+\kappa_{sr})+b\bar{\gamma}_{sr}}{\mu_{sr}(1+\kappa_{sr})}t.$$ (A.18)

The above substitution converts the upper limit of the integral to unity without
changing the lower limit of the integration. This makes it easy to represent
this integral into standard form of confluent Lauricella's hypergeometric function of three variables. It is defined as

$$\Phi_1^{(3)}(a,b_1,b_2,c,x,y,z) = \frac{\Gamma(c)}{\Gamma(a)\Gamma(c-a)}\int_0^1 t^{a-1}(1-t)^{c-a-1}(1-xt)^{-b_1}(1-yt)^{-b_2}\exp(zt)dt,$$ (A.19)

Further simplifications after substitution of (A.18) into (A.17) brings the integral in the form that can be given as

$$I_2 = \frac{c}{\pi} \frac{\sqrt{b\bar{\gamma}_{sr}}(\mu_{sr}\kappa_{sr})^{\mu_{sr}}}{2} \left(\frac{\mu_{sr}(1+\kappa_{sr})}{\mu_{sr}(1+\kappa_{sr})+b\bar{\gamma}_{sr}}\right)^{\mu_{sr}+1/2} \times$$

$$\int_0^1 \frac{y^{\mu_{sr}-1/2} \exp(\frac{\mu_{sr}(1+\kappa_{sr})}{\mu_{sr}(1+\kappa_{sr})+b\bar{\gamma}_{sr}} y)(1 - \frac{2\mu_{sr}(1+\kappa_{sr})+b\bar{\gamma}_{sr}}{2\mu_{sr}(1+\kappa_{sr})+2b\bar{\gamma}_{sr}} y)^{-1/2}}{\sqrt{2\mu_{sr}(1+\kappa_{sr})}(1 - \frac{\mu_{sr}(1+\kappa_{sr})}{\mu_{sr}(1+\kappa_{sr})+b\bar{\gamma}_{sr}} y)} dy. \quad (A.20)$$

The above expression can be compared with the definition of confluent Lauricella's hypergeometric function given in (A.19) to obtain the arguments of the function as

$$a = \mu_{sr} + 1/2, \quad (A.21)$$

$$b_1 = 1, \quad (A.22)$$

$$b_2 = 1/2, \quad (A.23)$$

$$c = \mu_{sr} + 3/2, \quad (A.24)$$

$$x = \frac{\mu_{sr}(1+\kappa_{sr})}{\mu_{sr}(1+\kappa_{sr})+b\bar{\gamma}_{sr}}, \quad (A.25)$$

$$y = \frac{2\mu_{sr}(1+\kappa_{sr})}{2\mu_{sr}(1+\kappa_{sr})+2b\bar{\gamma}_{sr}}, \quad (A.26)$$

$$z = \frac{\mu_{sr}(1+\kappa_{sr})}{\mu_{sr}(1+\kappa_{sr})+b\bar{\gamma}_{sr}}, \quad (A.27)$$

Thus, $I_2$ can be finally evaluated in the form of confluent Lauricella's hypergeometric function as

$$I_2 = \frac{c}{\pi} \frac{\sqrt{b\bar{\gamma}_{sr}}(\mu_{sr}\kappa_{sr})^{\mu_{sr}}}{2\sqrt{2\mu_{sr}(1+\kappa_{sr})}} \left(\frac{\mu_{sr}(1+\kappa_{sr})}{\mu_{sr}(1+\kappa_{sr})+b\bar{\gamma}_{sr}}\right)^{\mu+1/2} \frac{\Gamma(\mu_{sr}+1/2)\sqrt{\pi}}{\Gamma(\mu_{sr}+1)} \times$$

$$\Phi_1^{(3)}\left(\mu_{sr}+1/2,1,1/2,3/2,\frac{\mu_{sr}(1+\kappa_{sr})}{\mu_{sr}(1+\kappa_{sr})+b\bar{\gamma}_{sr}},\frac{2\mu_{sr}(1+\kappa_{sr})}{2\mu_{sr}(1+\kappa_{sr})+2b\bar{\gamma}_{sr}},\frac{\mu_{sr}(1+\kappa_{sr})}{\mu_{sr}(1+\kappa_{sr})+b\bar{\gamma}_{sr}}\right), \quad (A.28)$$

Note that $I_1$ & $I_2$ are respectively in the form of confluent hypergeometric function and confluent Lauricella's function. These functions are not commonly available in the mathematical computation softwares. Thus, numerical evaluation methods for finite integrals may be used. Alternatively, these functions can be numerically evaluated using their series representation. The confluent hypergeometric function in series form can be given as

$$\Phi_1(a,b,c,x,y) = \sum_{J=0}^{\infty}\sum_{n=0}^{\infty} \frac{(a)_{J+n}(b)_J x^J y^n}{(c)_{J+n} J!n!}, \quad (A.29)$$

The condition for convergence of this function is $|x| < 1$, which is satisfied in our case for all values of average SNR. The confluent Lauricella's hypergeometric function in series form can be given as

$$\Phi_1^{(3)}(a,b,c,x,y,z) = \sum_{J=0}^{\infty}\sum_{n=0}^{\infty}\sum_{p=0}^{\infty} \frac{(a)_{J+n+p}(b_1)_J (b_2)_n x^J y^n z^p}{(c)_{J+n+p} J!n!p!}. \quad (A.30)$$

Using (A.30) into (A.28), the integral $I_2$ can be expressed as,

$$I_2 = \frac{c\sqrt{b\bar{\gamma}_{sr}}\left(\mu_{sr}N_S N_R \kappa_{sr}\right)^{\mu_{sr}N_S N_R}}{2\pi\sqrt{2\mu_{sr}N_S N_R(1+\kappa_{sr})}} \left(\frac{\mu_{sr}N_S N_R(1+\kappa_{sr})}{\mu_{sr}N_S N_R(1+\kappa_{sr})+b\bar{\gamma}_{sr}}\right)^{\mu_{sr}N_S N_R + \frac{1}{2}} \frac{\Gamma\left(\mu_{sr}N_S N_R + \frac{1}{2}\right)\sqrt{\pi}}{\Gamma(\mu_{sr}N_S N_R + 1)} \times$$

$$\sum_{J=0}^{\infty}\sum_{n=0}^{\infty}\sum_{p=0}^{\infty} \frac{\left\{\begin{array}{l}\left(\mu_{sr}N_S N_R + \frac{1}{2}\right)_{J+n+p}(1)_J\left(\frac{1}{2}\right)_n\left(\frac{\mu_{sr}N_S N_R(1+\kappa_{sr})}{\mu_{sr}N_S N_R(1+\kappa_{sr})+b\bar{\gamma}_{sr}}\right)^J \times \\ \left(\frac{2\mu_{sr}N_S N_R(1+\kappa_{sr})}{2\mu_{sr}N_S N_R(1+\kappa_{sr})+2b\bar{\gamma}_{sr}}\right)^n \left(\frac{\mu_{sr}N_S N_R(1+\kappa_{sr})}{\mu_{sr}N_S N_R(1+\kappa_{sr})+b\bar{\gamma}_{sr}}\right)^n\end{array}\right\}}{\left(\frac{3}{2}\right)_{J+n+p} J!n!p!}. \quad (A.31)$$

# Authors

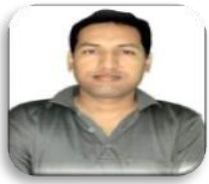
Ravi Shankar received his B.E. degree in Electronics and Communication Engineering from Jiwaji University, Gwalior, India, in 2006. He received his M.Tech. degree in Electronic and Communication Engineering from GGSIPU, New Delhi, India, in 2012. He received PhD in Wireless Communication from National Institute of Technology Patna, Patna, India, in 2019. He was an Assistant Professor at MRCE Faridabad, from 2013 to 2014, and has been engaged in researching in wireless communication networks. He is presently an Assistant Professor at MITS Madanapalle, Madanapalle, India. His current research interests are Cooperative Communication, D2D Communication, IoT/M2M networks, and network protocols for the networks. He is a student member of IEEE.
E-mail: ravishankar@mits.ac.in
ECE Department
MITS Madanapalle
Madanapalle-517325, AP, India

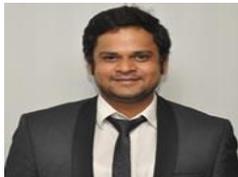
Lokesh Bhardwaj received the B. Tech degree in Electronics and Communication Engineering from Maharishi Dayanand University, Rohtak in 2007 and M. Tech in Electronics and Communication Engineering from Thapar University, Patiala in 2010. He is currently working as Assistant Professor with the Department of Electronics and Communication Engineering, Manav Rachna University, Faridabad. His research interest includes Wireless Communication. He is currently pursuing Ph.D from National Institute of Technology, Patna under the supervision of Dr. Ritesh Kumar Mishra in the area of "MIMO signal processing". His current research interests are Massive MIMO, D2D Communication, M2M networks, and network protocols for the networks. He is a student member of IEEE.
E-mail: lokesh@mru.edu.in
ECE Department
National Institute of Technology Patna
Patna-800005, India

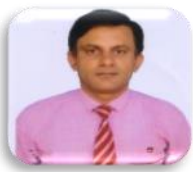
R K Mishra received his B.E. degree in Electronics and Communication Engineering from Shivaji University, Kolhapur, India, in 1998. He received his M.Tech. degree in Electronic and Communication Engineering from University of Burdwan, WB, India, in 2004. He received PhD in Wireless Communication from LNMU, Bihar, India, in 2011. He is presently working as an Assistant Professor at National Institute of Technology Patna, Patna, India and has been engaged in researching in wireless communication networks. His current research interests are Cooperative Communication, D2D Communication, IoT/M2M networks, and network protocols for the networks. He is a member of IEEE, Institution of Engineers and Indian Society of Technical Education.

E-mail: ritesh@nitp.ac.in  
ECE Department  
National Institute of Technology Patna  
Patna-800005, India